\documentclass[aps,pra,superscriptaddress,preprint]{revtex4-1}
\usepackage{subfigure}
\usepackage{xcolor}
\usepackage{graphicx}
\usepackage{bm}
\usepackage{amsmath}
\usepackage{float}
\usepackage{ulem}
\usepackage[colorlinks,linkcolor=blue,anchorcolor=blue,citecolor=blue,urlcolor=blue]{hyperref}

\begin{document}
\title{Detecting Collective Excitations in Self-Gravitating Bose-Einstein Condensates via Faraday Waves}
\author{Ning Liu}
\email{ningliu@mail.bnu.edu.cn}
\affiliation{School of Mathematics and Physics, Anqing Normal University, Anqing 246133, China}
\affiliation{Institute of Astronomy and Astrophysics, Anqing Normal University, Anqing 246133, China}
\affiliation{Key Laboratory of Multiscale Spin Physics(Ministry of Education), Beijing Normal University, Beijing 100875, China}
\author{Guodong Cheng}
\affiliation{School of Physics and Astronomy, Beijing Normal University, Beijing 100875, China}
\date{\today}

\begin{abstract}
We propose Faraday waves as a probe for collective excitations in self-gravitating Bose-Einstein condensates (SGBECs). Using a semi-classical approach based on linear stability analysis of the Gross-Pitaevskii-Newton equations, we derive a damped Mathieu equation governing parametric instabilities. Our analysis reveals well-separated regions of parametric resonance and Jeans instability in parameter space, with distinct growth rate characteristics: Jeans instability decreases monotonically to zero at the critical wavenumber $k_J$, while parametric resonance exhibits non-monotonic behavior with a clear maximum. These findings provide explicit experimental guidelines for accessing the parametric resonance regime. Numerical simulations demonstrate the transition from Faraday wave formation to Jeans collapse as gravitational strength increases, validating our theoretical framework.
\end{abstract}

\maketitle

\section{Introduction}
Gravity, the first fundamental interaction to be characterized, remains the least understood in quantum regimes. Persistent uncertainty surrounds its fundamental nature, specifically the question of its quantum or classical character~\cite{oppenheim2023postquantum,kryhin2025distinguishable}. This uncertainty stems from the extreme weakness of gravitational interactions, which prevents direct experimental probes of quantum gravity. Analog quantum systems provide a valuable alternative by emulating gravitational phenomena through engineered long-range attractive interactions. Although unable to directly test the quantum nature of gravity itself, these systems enable the study of quantum many-body behavior under gravitational-like potentials. They offer insights into collective quantum dynamics that may parallel those in real gravitational systems~\cite{barcelo2011analogue,liang2024evidence,marletto2025quantum}.

Within this context, self-gravitating quantum many-body systems have attracted growing interest as analog platforms for exploring gravitational phenomena in controllable quantum settings~\cite{chavanis2002phase,andersson2009static,grossardt2016approximations,ourabah2020quasiequilibrium,ourabah2023collective,asakawa2024collective}. These systems are fundamentally important for testing quantum-gravitational phenomena and have wide-ranging implications for modeling astrophysical scenarios. However, even foundational issues remain contentious. The form of the collective excitation spectrum, for instance, is debated. A hydrodynamic approach yields a classical-like spectrum~\cite{chavanis2017dissipative}, while a direct perturbative treatment of the quantum equations predicts ultraviolet divergence~\cite{akbari2024divergent}. This discrepancy highlights fundamental questions about the appropriate theoretical framework for describing collective excitations in self-gravitating quantum systems. Resolving these theoretical discrepancies requires not only refined theoretical approaches but also experimental access to the excitation spectrum of such systems.

Ultracold atomic gases offer an ideal platform for quantum simulation of these phenomena due to their exceptional purity and tunability~\cite{morsch2006dynamics,bloch2008many}. Through techniques such as optical dipole forces or laser-induced interactions, which generate long-range attractive interactions analogous to gravity~\cite{o2000bose,giovanazzi2001self,choi2002collision,chalony2013long}, they enable the emulation of self-gravitating quantum fluids and the study of collective behavior under controlled, gravitational-like potentials.

Building on these experimental capabilities, we focus specifically on self-gravitating Bose-Einstein condensates (SGBECs). SGBECs are governed by the Gross-Pitaevskii-Newton (GPN) equations, which couple the nonlinear Schr\"{o}dinger equation describing the condensate wavefunction to the Poisson equation for the self-consistent gravitational potential~\cite{paredes2020optics}. The GPN framework has enabled extensive analysis of SGBECs, including nonlinear dynamics~\cite{cartarius2008dynamics}, equilibrium configurations~\cite{guzman2006gravitational}, and quantized vortex formation~\cite{asakawa2024corotation}. Beyond their role as laboratory analogs, SGBECs are of significant interest in astrophysics and cosmology~\cite{hu2000fuzzy,boehmer2007can,jones2001theself,schroven2015stability,chavanis2020jeans,guo2021two,kirejczyk2022self,shukla2024gravity,sivakumar2025revealing}. They serve as models for ultralight bosonic dark matter candidates~\cite{chavanis2011mass,chavanis2011mass2,chavanis2014self,suarez2014review,verma2021formation,chavanis2021jeans,nikolaieva2021stable,nikolaieva2023stable}, potentially explaining core structures in galactic halos, and for boson stars~\cite{kolb1993axion,chavanis2012bose,dmitriev2021instability,verma2022rotating,levkov2017relativistic,dmitriev2024self}, hypothetical compact objects composed of scalar fields. Understanding the dynamics and collective excitations of SGBECs is thus crucial for probing fundamental physics across scales.

To study collective excitations in SGBECs, Faraday waves offer a powerful diagnostic tool~\cite{hernandez2021faraday,cominotti2022observation}. These waves are striking standing wave patterns that form on a fluid's surface when an external control parameter varies periodically. This process, known as parametric driving, was first documented by Michael Faraday in 1831~\cite{faraday1831xvii}. A common example is the vertical oscillation of the system. Faraday waves represent a spontaneous breaking of both temporal and spatial symmetries. Significantly, Faraday waves have been investigated in diverse quantum systems~\cite{staliunas2002faraday,staliunas2004faraday,nicolin2007faraday,bhattacherjee2008faraday,nath2010faraday,nicolin2010faraday,pollack2010collective,nicolin2011resonant,capuzzi2011suppression,balavz2012faraday,lakomy2012faraday,raportaru2012formation,balavz2014faraday,abdullaev2015faraday,sudharsan2016faraday,nguyen2019parametric,vudragovic2019faraday,abdullaev2019faraday,abdullaev2019faraday,turmanov2020generation,liu2020dynamics,cheng2021many,zhang2020pattern,maity2020parametrically,kwon2021spontaneous,zhang2022faraday,shukuno2023faraday,brito2023faraday,liang2024spin,wan2024spatiotemporal,del2024spatial,wang2025parametric}, including their first experimental demonstration in atomic BECs by Engels et al.~\cite{engels2007observation} and subsequent studies in dipolar condensates~\cite{lakomy2012faraday,turmanov2020generation}, two-component BECs~\cite{chen2019faraday,wang2025parametric}, spin-orbit-coupled BECs~\cite{zhang2022faraday} and Rydberg-dressed BECs~\cite{shi2023faraday}. These theoretical analyses and experiments establish that Faraday waves directly probe the underlying collective excitation spectrum of the quantum fluid. Therefore, driving Faraday waves in SGBECs offers a novel and controlled framework for investigating the collective dynamics, particularly the excitation spectrum of quantum fluids dominated by long-range attractive interactions.

In this paper, we present a theoretical framework for Faraday waves in SGBECs, driven by periodic modulation of the $s$-wave scattering length. We employ linear stability analysis of the GPN equations, decomposing perturbations into real and imaginary components within the local density approximation. This semi-classical approach provides a tractable framework for analyzing parametric instabilities while retaining essential quantum features through the quantum pressure term. Our analysis yields a damped Mathieu equation that governs the parametric instability, enabling us to construct detailed stability phase diagrams that reveal the distinct regions of parametric resonance and Jeans instability in parameter space. We systematically characterize the growth rates of both instabilities and analyze the dependence of the Faraday wavevector on driving parameters and gravitational strength. Furthermore, we incorporate dissipation effects and present numerical simulations that illustrate the formation and evolution of Faraday wave patterns. Our approach bridges quantum and classical descriptions, offering insights into the nature of collective excitations in self-gravitating quantum systems.

The paper is structured as follows: Section \ref{ML} outlines the theoretical model and introduces the GPN equation with time-dependent interaction strength. It then details the linear stability analysis, which leads to a Mathieu-like equation. Section \ref{SP} presents the resulting stability phase diagrams, analyzes the characteristics of the Faraday wavevector, and provides experimental parameter selection guidelines. Section \ref{NR} discusses numerical simulations of Faraday wave dynamics within the driven SGBEC, including the effects of dissipation. We conclude in Section \ref{C}.

\section{Model and Linear Stability Analysis}\label{ML}
\subsection{The Gross-Pitaevskii-Poisson Equations}

The dynamics of SGBECs is governed by the coupled Gross-Pitaevskii-Poisson (GPP) equations:
\begin{subequations}\label{gpp}
\begin{align}
\mathrm{i}\hbar \frac{\partial \Psi}{\partial t}
  &= -\frac{\hbar^2}{2m}\nabla^2\Psi
     + V_{\mathrm{ext}}\Psi
     + g(t)|\Psi|^2\Psi
     + m\Phi\Psi,   \label{gpp:a}\\[4pt]
\nabla^2\Phi
  &= 4\pi G m|\Psi|^2, \label{gpp:b}
\end{align}
\end{subequations}
where $\Psi(\mathbf{r},t)$ is the condensate wavefunction, $V_{\rm ext}$ denotes the external potential, $m$ is the particle mass, and $\Phi(\mathbf{r},t)$ is the gravitational potential satisfying the Poisson equation. The nonlinear coupling strength $g(t) = 4\pi\hbar^2 a_s(t)/m$ depends on the $s$-wave scattering length $a_s(t)$, which can be precisely controlled via Feshbach resonance techniques.

The GPP system admits an equivalent integral formulation:
\begin{equation}
\begin{aligned}
{\rm i}\hbar \frac{\partial \Psi}{\partial t} = &-\frac{\hbar^2}{2m}\nabla^2\Psi + V_{\rm ext}\Psi + g(t)|\Psi|^2\Psi \\
&- m^2G \int {\rm d}\mathbf{r}' \frac{|\Psi(\mathbf{r}',t)|^2}{|\mathbf{r}-\mathbf{r}'|}\Psi.
\end{aligned}
\end{equation}
The apparent singularity in the integral kernel is regularized by the finite extent of the wavefunction $|\Psi(\mathbf{r}',t)|^2$, which provides a natural cutoff. For physical wavefunctions with finite spatial support, the integral remains well-defined. The equivalence between differential and integral formulations can be rigorously established using distribution theory, where the Newtonian potential $-1/|\mathbf{r}-\mathbf{r}'|$ serves as the fundamental solution of the Laplacian operator.

In the absence of gravitational interactions ($\Phi=0$), Eq.~(\ref{gpp}) reduces to the standard Gross-Pitaevskii equation describing conventional BECs. Conversely, when mean-field interactions are neglected ($g=0$), one obtains the Schr\"{o}dinger-Newton equation, which has been extensively studied in contexts ranging from quantum measurement theories to models of dark matter.

A generalized formulation incorporating both logarithmic nonlinearity and dissipation has been proposed~\cite{chavanis2017dissipative}:
\begin{equation}
\begin{aligned}
{\rm i}\hbar \frac{\partial \Psi}{\partial t} = &-\frac{\hbar^2}{2m}\nabla^2\Psi + V_{\rm ext}\Psi + g|\Psi|^2\Psi + m\Phi\Psi \\
&-\beta^{-1}\ln |\Psi| \Psi - {\rm i}\hbar\gamma\Psi,
\end{aligned}
\end{equation}
where the logarithmic term $-\beta^{-1}\ln|\Psi|\Psi$ accounts for finite-temperature effects, with $\beta^{-1} = -k_B T$, where $k_B$ is Boltzmann's constant and $T$ denotes temperature. The dissipative term $-{\rm i}\hbar\gamma\Psi$ phenomenologically models energy losses, with $\gamma$ representing the damping coefficient. The original GPP system is recovered in the limits $\beta^{-1} \to 0$ and $\gamma \to 0$.

\subsection{Equilibrium State and Parametric Driving}

For self-gravitating Bose-Einstein condensates, the equilibrium configuration naturally assumes a Gaussian profile due to the balance between quantum pressure, contact interactions, and gravitational attraction. The ground-state wavefunction takes the form:
\begin{equation}
\Psi_0(\mathbf{r},t) = \sqrt{n_0(\mathbf{r})} e^{-i\mu t/\hbar}, \label{gaussian_background}
\end{equation}
with the density distribution:
\begin{equation}
n_0(\mathbf{r}) = n_{\text{max}} \exp\left(-\frac{r^2}{2\sigma^2}\right). \label{gaussian_density}
\end{equation}
Here $\sigma$ characterizes the spatial extent of the condensate, $n_{\text{max}}$ denotes the peak density, and $\mu$ is the chemical potential. The corresponding gravitational potential $\Phi_0$ satisfies the self-consistent Poisson equation $\nabla^2\Phi_0 = 4\pi G m n_0(\mathbf{r})$.

Faraday waves in Bose-Einstein condensates can be excited through two primary methods: periodic modulation of the external trapping potential or time-dependent variation of the interaction strength via Feshbach resonance~\cite{engels2007observation}. This work employs the latter approach, modulating the scattering length as
$g(t) = g_0[1 + \alpha\cos(\omega t)]$, where $g_0 = 4\pi\hbar^2 a_s/m$ is the unmodulated coupling constant, $\omega$ represents the driving frequency, and $\alpha$ quantifies the modulation depth. This parametric driving scheme provides a controlled mechanism for exciting collective modes in the quantum fluid.

\subsection{Linear Stability Analysis}

To investigate the emergence of Faraday waves, we introduce small perturbations around the equilibrium state:
\begin{equation}
\Psi(\mathbf{r},t) = \Psi_0(\mathbf{r},t)\left[1 + \delta\psi(\mathbf{r},t)\right]. \label{local_perturb}
\end{equation}
For perturbations with wavelengths much shorter than the system size ($k\sigma \gg 1$), we employ the local density approximation (LDA). In this regime, the background density $n_0(\mathbf{r})$ varies slowly compared to the rapid oscillations of the perturbation, allowing us to treat the system as locally homogeneous. The perturbation takes the plane-wave form:
\begin{equation}
\delta\psi(\mathbf{r},t) = [u(t) + iv(t)]e^{i\mathbf{k}\cdot\mathbf{r}}, \label{plane_wave_perturb}
\end{equation}
where $u(t)$ and $v(t)$ represent the real and imaginary parts of the perturbation amplitude, respectively. The corresponding density perturbation $\delta n(\mathbf{r},t) = |\Psi|^2 - |\Psi_0|^2$ becomes:
\begin{equation}
\delta n(\mathbf{r},t)= 2n_0(\mathbf{r})u(t)\cos(\mathbf{k}\cdot\mathbf{r}).
\end{equation}
Due to the rapid oscillations of the plane-wave perturbation over the Gaussian background, we employ spatial averaging, defining the spatially averaged density:
\begin{equation}
\bar{n}_0 = \frac{1}{V}\int n_0(\mathbf{r}) {\rm d}^3r = \frac{n_{\text{max}}}{(2\pi)^{3/2}\sigma^3}. \label{avg_density}
\end{equation}
This averaging procedure is physically motivated by the fact that the perturbation effectively samples the entire condensate volume, and the rapid oscillations cause gradient terms to average to zero.

The gravitational potential perturbation satisfies:
\begin{equation}
\nabla^2(\delta\Phi) = 4\pi G m \delta n = 8\pi G m \bar{n}_0 u(t)\cos(\mathbf{k}\cdot\mathbf{r}).
\end{equation}
Under the LDA, where terms involving gradients of the density can be neglected compared to the dominant $-k^2 \cos(\mathbf{k}\cdot\mathbf{r})$ term, we obtain the particular solution:
\begin{equation}
\delta\Phi(\mathbf{r},t) = -\frac{8\pi G m \bar{n}_0 u(t)}{k^2}\cos(\mathbf{k}\cdot\mathbf{r}).
\end{equation}
This approximation is valid when $k\sigma \gg 1$, ensuring that the perturbation wavelength is much smaller than the characteristic size of the condensate.

Substituting the perturbed wavefunction into the GPP equations with dissipation and retaining terms linear in the perturbation yields coupled equations for $u(t)$ and $v(t)$. After spatial averaging over the Gaussian profile, we obtain:
\begin{subequations}
\begin{align}
-\hbar\frac{{\rm d} v}{{\rm d} t} &= \frac{\hbar^2 k^2}{2m}u + [g_0 + 2g_0\alpha\cos(\omega t)]\bar{n}_0 u - \frac{4\pi G m^2 \bar{n}_0 u}{k^2} + \hbar\gamma v, \label{realm_eq} \\
\hbar\frac{{\rm d} u}{{\rm d} t} &= \frac{\hbar^2 k^2}{2m}v + [2g_0\alpha\cos(\omega t) + g_0]\bar{n}_0 v - \hbar\gamma u, \label{imagm_eq}
\end{align}
\end{subequations}
where $\bar{n}_0$ is the spatially averaged density defined in Eq.~(\ref{avg_density}). 

Substituting Eq.~(\ref{imagm_eq}) into Eq.~(\ref{realm_eq}) and applying the adiabatic approximation—neglecting time derivatives of slowly varying coefficients when the modulation frequency $\omega$ is much smaller than the system's characteristic frequency—we obtain the second-order equation (The complete derivation of the linearized equations, including detailed treatment of all terms in the GPP equations, is provided in Appendix \ref{app:detailed_derivation}.):
\begin{equation}
\frac{{\rm d}^2 u}{{\rm d} t^2} + 2\gamma\frac{{\rm d} u}{{\rm d} t} + \left[\frac{\hbar^2 k^4}{4m^2} + \frac{g(t)\bar{n}_0 k^2}{m} - \Omega_J^2\right]u = 0, \label{second_order}
\end{equation}
Equation (\ref{second_order}) reveals three distinct physical contributions: the quantum pressure term $\hbar^2 k^4/4m^2$ arising from kinetic energy, the interaction term $g(t)\bar{n}_0 k^2/m$ representing mean-field effects, and the gravitational term $\Omega_J^2 = 4\pi G m \bar{n}_0$, which encodes the Jeans instability. The Jeans frequency derives its name from Sir James Jeans, who in 1902 first analyzed the gravitational instability of an infinite homogeneous medium in the context of astrophysics~\cite{jeans1902stability}. The Jeans criterion determines when self-gravity overcomes pressure support, leading to collapse~\cite{khlopov1985gravitational}. In our quantum analog, $\Omega_J$ represents the characteristic frequency at which self-gravity operates, playing a role analogous to the classical Jeans frequency in astrophysical systems.

Substituting the modulated interaction strength $g(t) = g_0[1 + \alpha\cos(\omega t)]$ and introducing dimensionless time $\tau = \omega t/2$, Eq.~(\ref{second_order}) yields the damped Mathieu equation:
\begin{equation}
\frac{{\rm d}^2 u}{{\rm d}\tau^2} + \bar{\gamma}\frac{{\rm d} u}{{\rm d}\tau} + \left[\lambda - 2q\cos(2\tau)\right]u = 0, \label{mathieu_eq}
\end{equation}
with dimensionless parameters:
\begin{equation}
\lambda = \frac{4\Omega_k^2}{\omega^2}, \quad q = -\frac{2\alpha g_0 \bar{n}_0 k^2}{m\omega^2}, \quad \bar{\gamma} = \frac{4\gamma}{\omega}. \label{mathieu_params}
\end{equation}
Here, $\Omega_k$ characterizes the frequency of collective excitations in SGBECs:
\begin{equation}
\Omega_k^2 = \Omega_{\rm Bog}^2 - \Omega_J^2, \label{omega_k}
\end{equation}
where $\Omega_{\rm Bog} = \sqrt{\hbar^2 k^4/(4m^2) + g_0\bar{n}_0 k^2/m}$ is the Bogoliubov excitation spectrum.

\begin{figure}[!ht]
\centering
\includegraphics[width=0.6\linewidth]{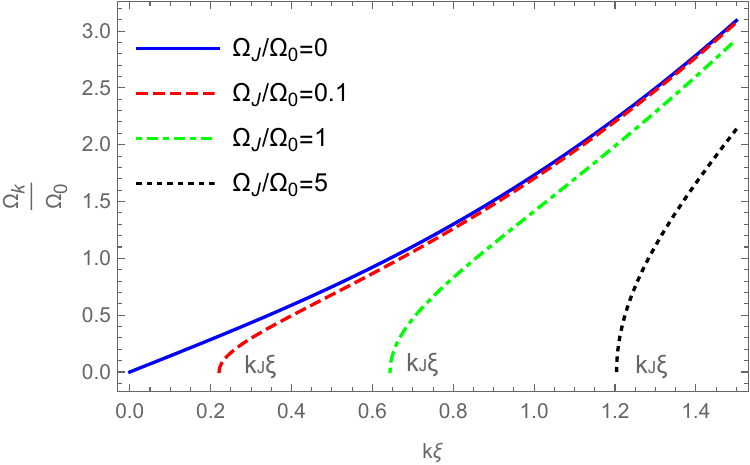}
\caption{(Color online) Collective excitation spectrum of SGBECs. Parameters: $\Omega_0 = g_0 \bar{n}_0/\hbar$, $\xi = \hbar/\sqrt{2 m g_0 \bar{n}_0}$. The vertical dashed line indicates $k_J$ where $\Omega_k^2 = 0$, marking the onset of Jeans instability. For $k < k_J$ (specifically, $k_J \xi = \sqrt{\sqrt{1+\Omega_J^2/\Omega_0^2}-1}$), $\Omega_k^2 < 0$ indicates exponential growth, defining the quantum Jeans scale~\cite{chavanis2020jeans}.}
\label{sge}
\end{figure}

Figure~\ref{sge} illustrates the collective excitation spectrum of SGBECs. For $k \xi \gg 1$ (where $\xi = \hbar/\sqrt{2 m g_0 \bar{n}_0}$ is the healing length), $\Omega_k$ approaches the Bogoliubov dispersion. The vertical dashed line indicates the critical wavenumber $k_J$ where $\Omega_k^2 = 0$, marking the onset of Jeans instability. For $k < k_J$, $\Omega_k^2 < 0$ indicates exponential growth, defining the quantum Jeans scale. The Jeans instability ($k < k_J$) and parametric resonance ($k > k_J$ with appropriate $\alpha$) occupy distinct regions in parameter space, separated by the critical wavenumber $k_J$. 

\section{Stability and Parametric Resonance}\label{SP}
\subsection{Stability Phase Diagram and Growth Rate Analysis}

According to Floquet theory~\cite{mclachlan1947theory,abramowitz1965handbook,wang1989special}, the solution of Eq.~\eqref{mathieu_eq} can be expressed as $u(\tau) = e^{\mu \tau} P(\tau)$, where $P(\tau)$ is a periodic function and $\mu$ is the complex Floquet exponent. The stability is determined by the real part of $\mu$: $\Re(\mu) > 0$ indicates exponential growth (instability), while $\Re(\mu) \leq 0$ signifies stability or decay. The Mathieu equation exhibits a series of instability tongues in the $(\lambda, q)$ parameter space, corresponding to resonance conditions $\lambda = \nu^2$ for $\nu = 1, 2, 3, \ldots$.

In our analysis, we focus primarily on the dominant $\nu = 1$ resonance tongue, as higher-order tongues ($\nu \geq 2$) occupy progressively narrower regions in parameter space and are therefore less experimentally accessible. For the dominant tongue, the resonance condition $\lambda = 1$, i.e., $\Omega_k = \omega/2$, leads to the frequency-matching relation:
\begin{equation}
\left( \frac{\omega}{2} \right)^2 = \Omega_{\rm Bog}^2 - \Omega_J^2,
\label{fc}
\end{equation}
This relation provides the fundamental criterion for Faraday wave excitation in SGBECs, requiring matching between the parametrically driven frequency and the gravitationally modified collective excitations.

\begin{figure}[!ht]
\centering
\includegraphics[width=0.95\textwidth]{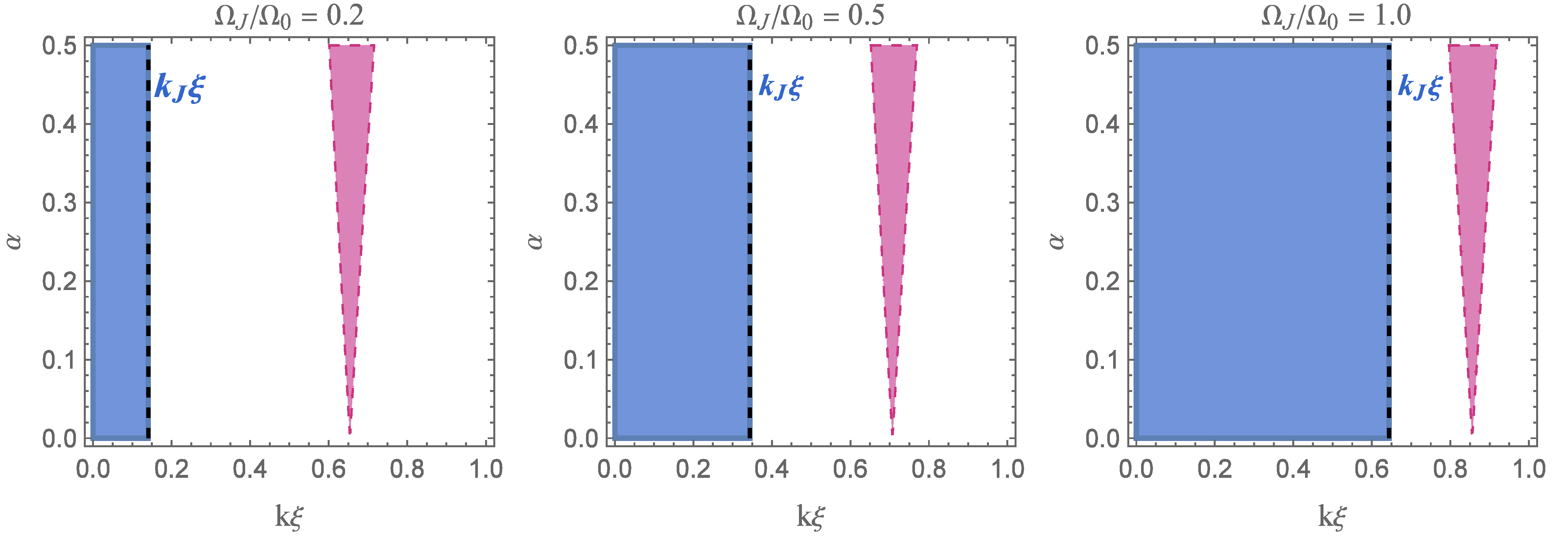}
\caption{(Color online) Stability phase diagrams in the $(k\xi, \alpha)$ parameter space for different gravitational strengths: (a) $\Omega_J/\Omega_0 = 0.2$, (b) $\Omega_J/\Omega_0 = 0.5$, and (c) $\Omega_J/\Omega_0 = 1.0$. Blue regions denote Jeans instability ($\lambda < 0$), while purple regions indicate parametric resonance within the first Mathieu instability tongue ($\lambda > 0$). The vertical dashed lines mark the critical Jeans wavenumber $k_J\xi$, which separates the two distinct instability regimes. Parameters: $\omega/\Omega_0 = 0.5$.}
\label{stability1}
\end{figure}

Figure~\ref{stability1} illustrates the distinct instability regimes in the parameter space $(k\xi, \alpha)$ for different gravitational strengths. The phase diagrams reveal a fundamental separation between Jeans instability and parametric resonance, with the critical Jeans wavenumber $k_J$ serving as the precise boundary between these mechanisms. As $\Omega_J/\Omega_0$ increases from 0.2 to 1.0, the Jeans instability region (blue) expands significantly due to enhanced gravitational attraction, while the parametric resonance tongue (purple) shifts toward larger wavenumbers. This systematic evolution demonstrates how gravitational strength fundamentally reshapes the stability landscape of driven SGBECs.

The distinct nature of these instability mechanisms is further elucidated by their characteristic growth rates. In the Jeans-unstable regime ($\lambda < 0$), density perturbations grow exponentially with time as $\delta n \propto e^{\Gamma_{\text{Jeans}} t}$, where the growth rate $\Gamma_{\text{Jeans}}$ quantifies how rapidly the gravitational instability develops. This growth rate is derived from the linear stability analysis of the GPP equations and takes the explicit form:
\begin{equation}
\Gamma_{\text{Jeans}} = \frac{2}{\omega}\sqrt{\Omega_J^2 - \left(\frac{\hbar^2 k^4}{4m^2} + \frac{g_0\bar{n}_0 k^2}{m}\right)}.
\label{jeans_growth}
\end{equation}

\begin{figure}[!ht]
\centering
\includegraphics[width=0.95\textwidth]{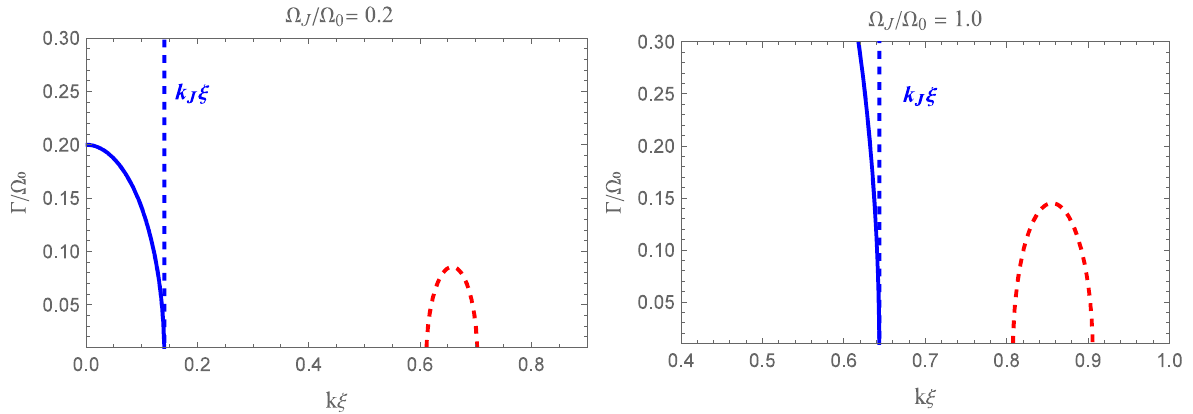}
\caption{(Color online) Comparison of instability growth rates for different gravitational strengths with fixed modulation depth $\alpha = 0.4$: (a) $\Omega_J/\Omega_0 = 0.2$ and (b) $\Omega_J/\Omega_0 = 1.0$. Blue solid curves: Jeans growth rate $\Gamma_{\text{Jeans}}/\Omega_0$; purple dashed curves: parametric resonance growth rate $\Gamma_{\text{parametric}}/\Omega_0$ obtained from numerical Floquet analysis. The vertical dashed lines indicate the critical Jeans wavenumber $k_J\xi$, where $\Gamma_{\text{Jeans}}$ vanishes as required by theory.}
\label{fig:growth_comparison}
\end{figure}

Mathematical analysis confirms the monotonic behavior of $\Gamma_{\text{Jeans}}(k)$. The derivative with respect to $k$ is given by:
\begin{equation}
\frac{d\Gamma_{\text{Jeans}}}{dk} = -\frac{1}{\omega\sqrt{f(k)}}\left(\frac{\hbar^2 k^3}{m^2} + \frac{2g_0\bar{n}_0 k}{m}\right),
\end{equation}
where $f(k) = \Omega_J^2 - \left(\frac{\hbar^2 k^4}{4m^2} + \frac{g_0\bar{n}_0 k^2}{m}\right) > 0$ for $0 < k < k_J$. Since the expression in parentheses is strictly positive for $k > 0$, the derivative is negative throughout the Jeans-unstable region. This rigorously establishes that $\Gamma_{\text{Jeans}}(k)$ decreases monotonically from its maximum value ${2\Omega_J}/{\omega}$ at $k = 0$ to zero at the critical Jeans wavenumber $k_J$, as shown in Fig.~\ref{fig:growth_comparison}. The vanishing growth rate at $k_J$ reflects the exact balance between gravitational collapse and stabilizing quantum pressure at the instability threshold.

In contrast, parametric resonance ($\lambda > 0$) displays an oscillatory growth pattern. Numerical analysis of the Mathieu equation reveals that the parametric resonance growth rate exhibits characteristic non-monotonic behavior in the $(k\xi, \Gamma)$ plane, reaching a maximum at intermediate wavenumbers before decreasing at large $k$, as shown in Fig.~\ref{fig:growth_comparison}. This complex behavior arises from the intricate structure of Mathieu instability tongues and reflects the competing effects of parametric driving and quantum pressure stabilization.

Figure~\ref{fig:growth_comparison} presents a quantitative comparison of these growth rates. For weak gravity ($\Omega_J/\Omega_0 = 0.2$), the Jeans instability region is confined to small wavenumbers with modest growth rates, while parametric resonance extends across a broad range of accessible parameter space. For strong gravity ($\Omega_J/\Omega_0 = 1.0$), the substantially larger Jeans instability region with enhanced growth rates necessitates careful parameter selection to access the parametric resonance regime for Faraday wave observation.

\begin{figure}[!ht]
\centering
\includegraphics[width=0.95\textwidth]{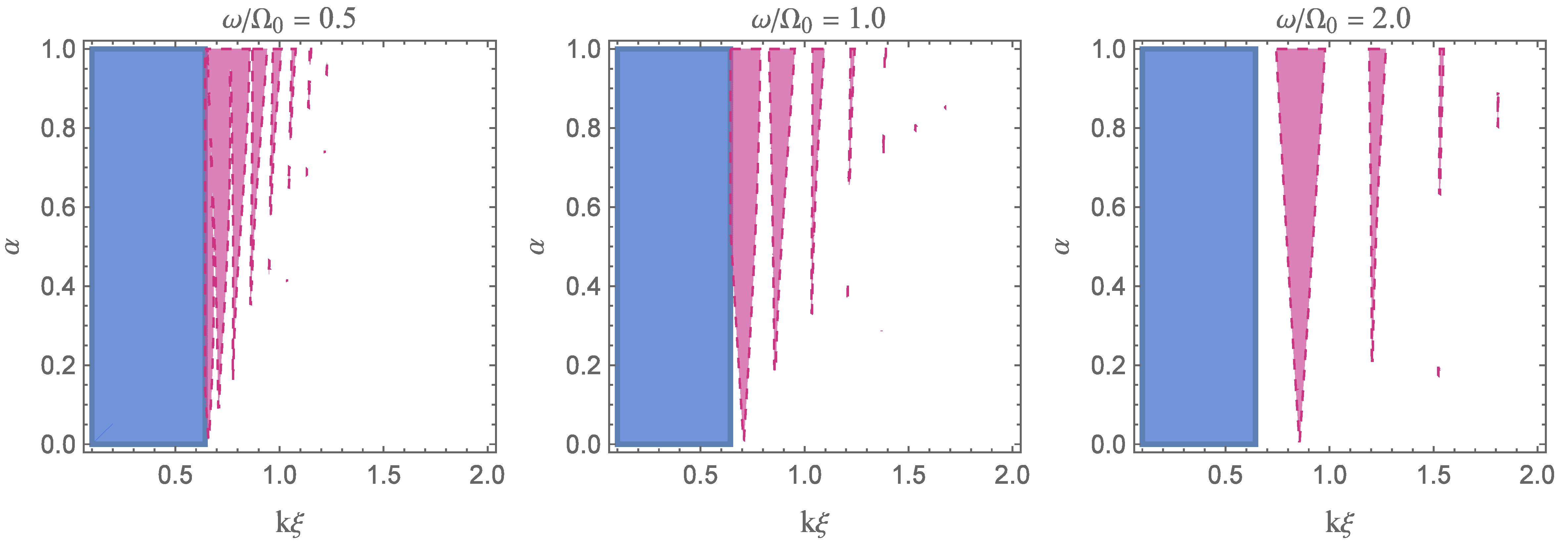}
\caption{(Color online) Stability phase diagrams at different driving frequencies with fixed gravitational strength $\Omega_J/\Omega_0 = 1.0$: (a) $\omega/\Omega_0 = 0.5$, (b) $\omega/\Omega_0 = 1.0$, and (c) $\omega/\Omega_0 = 2.0$. The parametric resonance tongue (purple) shifts systematically toward larger wavenumbers with increasing frequency, while the Jeans instability boundary (blue) remains fixed at $k_J$. This frequency control provides an experimental approach to optimize parametric resonance accessibility.}
\label{fig:frequency_evolution}
\end{figure}

The dependence of the stability phase diagram on driving frequency is shown in Fig.~\ref{fig:frequency_evolution}. While the Jeans instability boundary remains fixed at $k_J$ (determined solely by $\Omega_J/\Omega_0$), the parametric resonance tongue shifts systematically toward larger wavenumbers with increasing $\omega/\Omega_0$. This frequency dependence arises from the scaling $\lambda \propto \omega^{-2}$ and $q \propto \omega^{-2}$ in the Mathieu equation parameters. The ability to control the parametric resonance position relative to the fixed Jeans boundary through frequency selection represents a key experimental strategy.

\subsection{The Faraday Wavevector}

From the instability condition $\lambda = \nu^2$ for the $\nu^{\text{th}}$ resonance tongue, we obtain the Faraday wavevector $k_F$ that will be observed experimentally. In the zero-temperature limit ($\beta^{-1}=0$), its explicit expression is:
\begin{equation}
k_F \xi = \sqrt{ \sqrt{ 1 + \frac{1}{\Omega_0^2}\left( \Omega_J^2 + \frac{\nu^2 \omega^2}{4} \right) } - 1 }.\label{kfe_zeroT}
\end{equation}
In the absence of self-gravity ($\Omega_J=0$), this expression reduces to the well-known result for conventional BECs~\cite{hernandez2021faraday}:
\begin{equation}
k_F^{(0)} \xi = \sqrt{ \sqrt{1 + \left( \frac{\nu \omega}{2\Omega_0}\right)^2 } - 1 }.\label{kfe_standard}
\end{equation}
For this conventional case, the wavevector $k_F^{(0)}$ increases monotonically with the driving frequency $\omega$.

\begin{figure}[!htpb]
\centering
\includegraphics[width=0.55\linewidth]{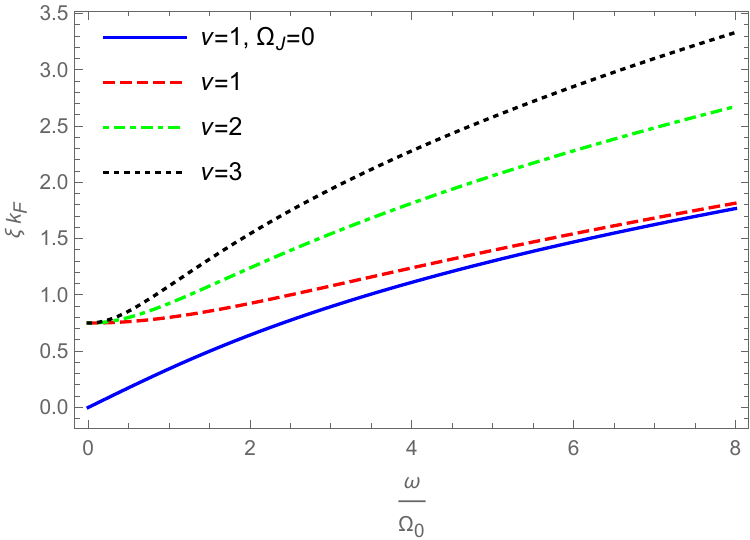}
\caption{(Color online) Faraday wavevector versus driving frequency for $\nu=1,2,3$ with $\Omega_J/\Omega_0 = 1.2$.}
\label{kf}
\end{figure}

Figure~\ref{kf} shows $k_F$ as a function of driving frequency $\omega$ for the first three branches. The Faraday wavevector increases with driving frequency, approaching the behavior of non-self-gravitating BECs at large $\omega$. At low driving frequencies, $k_F\xi$ approaches a finite value $\sqrt{\sqrt{\Omega_J^2/\Omega_0^2 + 1} - 1}$ in SGBECs, contrasting with the $k_F \rightarrow 0$ behavior in non-gravitational systems. Higher $\nu$ values yield larger $k_F$ at fixed $\omega$, consistent with excitation of shorter-wavelength modes.

\subsection{Experimental Parameter Selection Guidelines}

Based on our stability analysis, we provide specific guidelines for experimental parameter selection to successfully observe Faraday waves in SGBECs:

\begin{table}[!ht]
\centering
\caption{Experimental parameter selection guidelines for Faraday wave observation in SGBECs}
\label{tab:experimental_guidelines}
\begin{tabular}{p{3cm}p{4cm}p{6cm}}
\hline
\textbf{Parameter} & \textbf{Recommended Range} & \textbf{Physical Rationale} \\
\hline
$k\xi$ & $> k_J\xi + 0.2$ & Ensure operation within parametric resonance regime \\
\hline
$\alpha$ & 0.3--0.6 & Sufficient driving amplitude for parametric resonance \\
\hline
$\omega/\Omega_0$ & 1.0--2.0 & Optimize parametric resonance accessibility \\
\hline
$\Omega_J/\Omega_0$ & $<$ 0.8 & Limit Jeans instability region extent \\
\hline
\end{tabular}
\end{table}

The core experimental strategy emerging from our analysis is to select parameters that place the system securely within the parametric resonance tongue ($k > k_J$ with appropriate $\alpha$), ensuring that Faraday wave growth occurs on experimentally accessible time scales. The frequency-matching condition in Eq.~\eqref{fc} provides the theoretical foundation for this parameter selection. Our comprehensive stability analysis thus delivers practical experimental guidance while advancing the fundamental understanding of driven gravitational quantum fluids.

\section{Numerical Results}\label{NR}

Our numerical simulations solve the dimensionless damped Mathieu equation [Eq.~(\ref{mathieu_eq})] governing the perturbation amplitude $U(\tau)$. To investigate the parametric resonance regime, we select parameters corresponding to the primary resonance tongue: $\lambda = 1$ and $q = -0.1$, with $\bar{\gamma} \equiv 4\gamma/\omega$ representing the dimensionless damping coefficient. These parameter choices satisfy the frequency-matching condition $2\Omega_k = \omega$ for $\nu=1$ in the absence of dissipation.

\begin{figure}[!htpb]
\centering
\includegraphics[width=0.85\textwidth]{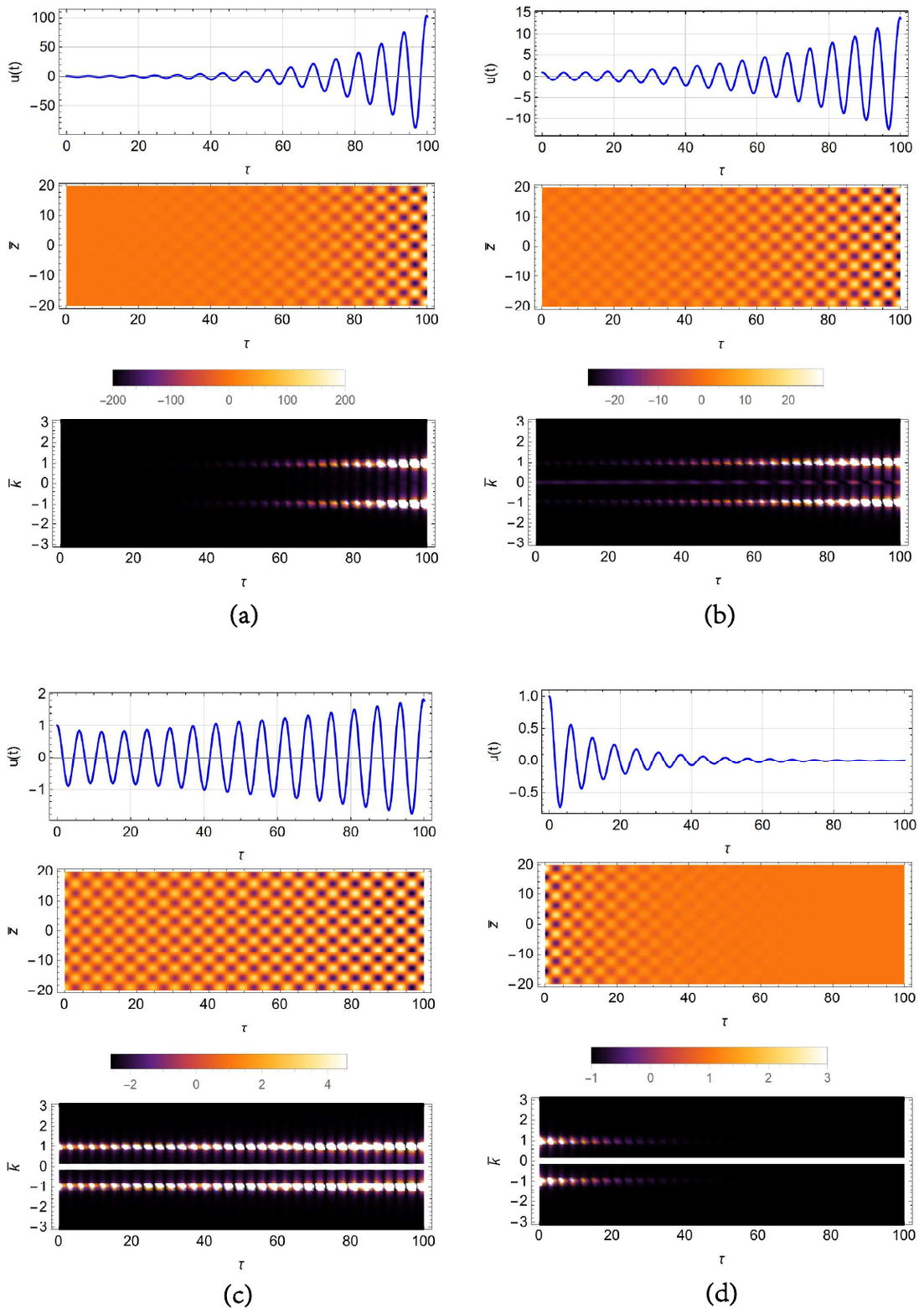}
\caption{(Color online) Density wave dynamics under different dissipation strengths with $\lambda=1$, $q=-0.1$: (a) $\bar{\gamma}=0$, (b) $\bar{\gamma}=0.01$, (c) $\bar{\gamma}=0.02$, (d) $\bar{\gamma}=0.05$. The panels from top to bottom show perturbation amplitude evolution, real-space density patterns, and momentum-space distributions, respectively.}
\label{kwo}
\end{figure}

Figure~\ref{kwo} presents the dynamical evolution across four dissipation strengths: (a) $\bar{\gamma}=0$ (dissipationless), (b) $\bar{\gamma}=0.01$ (weak damping), (c) $\bar{\gamma}=0.02$ (moderate damping), and (d) $\bar{\gamma}=0.05$ (strong damping). The top row displays the temporal evolution of $U(\tau)$, the middle row shows the background-subtracted density $|\Psi|^2/\bar{n}_0 - 1$ to isolate Faraday wave patterns from the uniform ground state, and the bottom row presents the momentum-space density distribution.

For $\bar{\gamma}=0$ [Fig.~\ref{kwo}(a)], parametric instability drives exponential growth of $U(\tau)$ with Floquet exponent $\mu \approx 0.2$, generating well-defined density waves with wavelength $\lambda_F = 2\pi/k$ and sharp momentum-space peaks at $\pm k$. At $\bar{\gamma}=0.01$ [Fig.~\ref{kwo}(b)], the system maintains exponential growth with minimal amplitude reduction, confirming that weak dissipation cannot suppress parametric resonance. At $\bar{\gamma}=0.02$ [Fig.~\ref{kwo}(c)], the combined effects of dissipation and parametric driving result in significantly suppressed amplitude growth and weak density modulations. Under strong dissipation ($\bar{\gamma}=0.05$, Fig.~\ref{kwo}(d)), dissipation dominates the dynamics, causing rapid decay of $U(\tau)$ within $\tau < 20$ and complete suppression of density waves. This sequence demonstrates the transition from parametric-resonance-dominated dynamics at low dissipation to dissipation-dominated behavior at high damping strengths.

\begin{figure}[!htpb]
\centering
\includegraphics[width=0.85\textwidth]{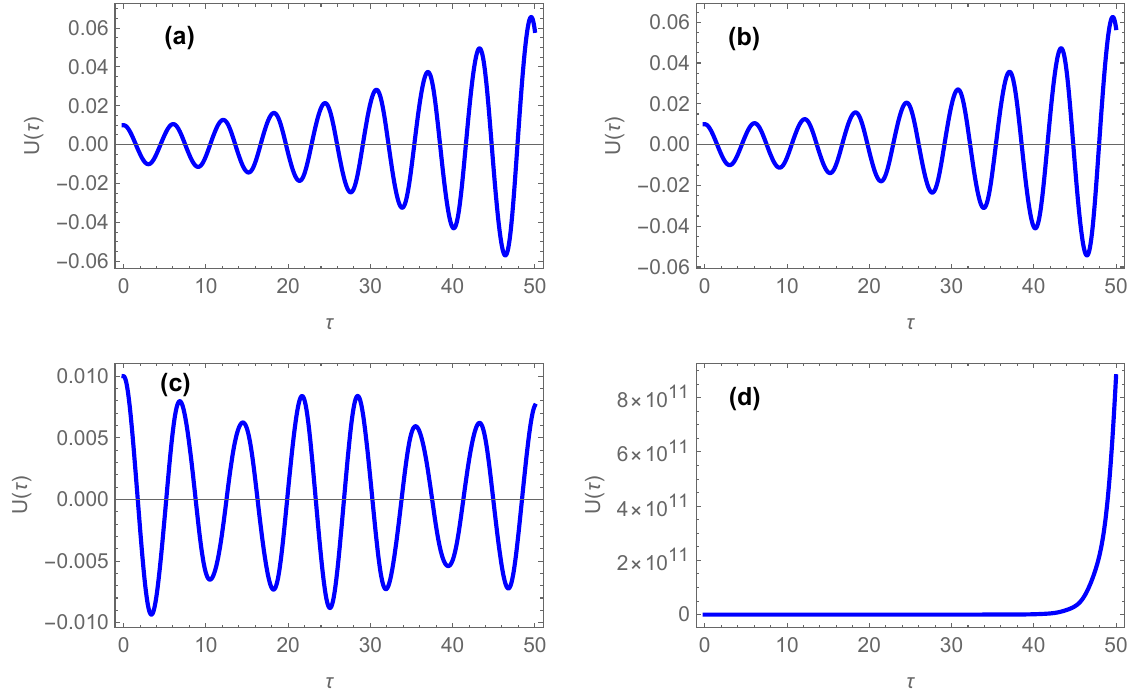}
\caption{(Color online) Evolution of perturbation amplitude $U(\tau)$ under varying gravitational strengths: (a) $\Omega_J/\Omega_0 = 0$ ($\lambda=1$), (b) $\Omega_J/\Omega_0 = 0.1$ ($\lambda=0.96$), (c) $\Omega_J/\Omega_0 = 0.5$ ($\lambda=0$), (d) $\Omega_J/\Omega_0 = 1.2$ ($\lambda=-4.76$). Fixed parameters: $\omega/\Omega_0 = 1$, $q = -0.1$, $\bar{\gamma} = 0.01$. The $\lambda$ values are determined by the resonance condition $\lambda = 4(\Omega_{\rm Bog}^2 - \Omega_J^2)/\omega^2$ with fixed $\Omega_{\rm Bog}$ corresponding to $\lambda=1$ at $\Omega_J=0$.}
\label{fwo}
\end{figure}

Figure~\ref{fwo} presents the evolution of perturbation amplitude $U(\tau)$ under varying Jeans frequencies, with fixed parameters $\omega/\Omega_0 = 1$, $q = -0.1$, and $\bar{\gamma} = 0.01$. The corresponding $\lambda$ values are determined by the relation $\lambda = 4(\Omega_{\rm Bog}^2 - \Omega_J^2)/\omega^2$, where $\Omega_{\rm Bog}$ is fixed to satisfy $\lambda=1$ at $\Omega_J=0$. This yields $\lambda = 1.00$, $0.96$, $0.00$, and $-4.76$ for $\Omega_J/\Omega_0 = 0$, $0.1$, $0.5$, and $1.2$, respectively.

In the absence of self-gravity ($\Omega_J/\Omega_0 = 0$, $\lambda = 1$; Fig.~\ref{fwo}a), the system resides at the center of the primary parametric resonance tongue, resulting in exponential growth of $U(\tau)$ characteristic of parametric instability that seeds Faraday waves. For weak gravity ($\Omega_J/\Omega_0 = 0.1$, $\lambda = 0.96$; Fig.~\ref{fwo}b), the system remains within the parametric resonance tongue but approaches its boundary, exhibiting sustained growth at a reduced rate with Faraday waves forming at diminished amplitudes. At the critical transition ($\Omega_J/\Omega_0 = 0.5$, $\lambda = 0$; Fig.~\ref{fwo}c) where $\Omega_k = 0$, the system exhibits irregular, non-exponential growth that deviates from characteristic Faraday patterns. In the Jeans-dominated regime ($\Omega_J/\Omega_0 = 1.2$, $\lambda = -4.76$; Fig.~\ref{fwo}d) where $\Omega_k^2 < 0$, self-gravity drives rapid exponential growth characteristic of gravitational collapse, suppressing Faraday wave formation.

This progression validates the stability phase diagram predictions, demonstrating how increasing gravitational strength systematically shifts the system from parametric resonance to Jeans instability dominance. The sensitivity of driven quantum fluids to long-range attractive interactions is highlighted by the sharp transition near $\Omega_J/\Omega_0 \sim 0.5$ ($\lambda = 0$), where the system behavior changes fundamentally.

\section{Conclusion}\label{C}
In this work, we have demonstrated that parametric resonance induced by periodic modulation of the $s$-wave scattering length can generate Faraday waves in self-gravitating Bose-Einstein condensates (SGBECs). Through linear stability analysis of the driven Gross-Pitaevskii-Newton equations, we derived a damped Mathieu equation governing the perturbation dynamics. Floquet analysis revealed a stability phase diagram that clearly separates parametric resonance and Jeans instability regions in parameter space, with the critical Jeans wavenumber $k_J$ providing a well-defined boundary between these distinct mechanisms. Growth rate analysis shows that the Jeans instability growth rate decreases monotonically from its maximum at $k=0$ to zero at $k_J$, while parametric resonance exhibits a non-monotonic behavior with an initial increase to a maximum followed by a decrease. The Faraday wavevector was systematically analyzed, with its dependence on gravitational strength characterized through comprehensive phase diagrams. Our analysis provides explicit experimental parameter selection guidelines to ensure clear observation of Faraday waves by operating in the parametric resonance regime. Numerical simulations illustrated the formation and evolution of density wave patterns under various dissipation conditions and Jeans frequencies; simulations reveal a characteristic transition from parametric-resonance-driven Faraday waves to gravity-dominated Jeans collapse as the Jeans frequency increases.

Our semi-classical approach, based on decomposing perturbations into real and imaginary components within the local density approximation, provides a tractable framework for analyzing parametric instabilities in self-gravitating quantum systems. While this method simplifies the full quantum dynamics, it retains essential quantum features through the quantum pressure term and yields a well-behaved excitation spectrum characterized by $\Omega_k^2 = \Omega_{\rm Bog}^2 - \Omega_J^2$. This result aligns more closely with the hydrodynamic approach that predicts a classical-like spectrum, rather than the ultraviolet divergence predicted by some direct perturbative quantum treatments. The absence of ultraviolet divergence in our results suggests that the semi-classical GPN framework provides a physically reasonable description of collective excitations in self-gravitating quantum fluids, at least in the long-wavelength regime where our approximations are valid. The Faraday wave method we propose offers an experimental pathway to probe this excitation spectrum and test the predictions of different theoretical approaches.

Notably, while this study employed modulation of the scattering length, Faraday waves may also be excited through periodic modulation of the external potential in similar configurations. Given the significance of the generalized GPN equations in astrophysical contexts, our approach provides a flexible platform for exploring gravitational analogs. Specifically, such systems could model scenarios where external potential fields originate from compact objects like black holes, and coupling constants emerge from self-gravitating dark matter configurations. This framework offers promising pathways for laboratory investigations into gravitational phenomena using quantum degenerate gases.

\paragraph*{Acknowledgments}
This work was supported by the Open Fund of Key Laboratory of Multiscale Spin Physics (Ministry of Education), Beijing Normal University (Grant No.SPIN2024N03), and the Scientific Research Startup Foundation for High-Level Talents at Anqing Normal University (Grant No.241042). We thank Professor Zhanchun Tu for valuable discussions and insightful comments.
\begin{appendix}
\section{Detailed Derivation of Linear Stability Analysis}
\label{app:detailed_derivation}

\subsection{Linearization of the GPN Equations}

We begin by substituting the perturbed wavefunction $\Psi = \Psi_0(1+\delta\psi)$ with $\delta\psi = (u+iv)e^{i\mathbf{k}\cdot\mathbf{r}}$ into the GPN equation. The time derivative term becomes:
\begin{align}
\mathrm{i}\hbar \frac{\partial \Psi}{\partial t} &= \mathrm{i}\hbar \frac{\partial}{\partial t}[\Psi_0(1+(u+iv)e^{i\mathbf{k}\cdot\mathbf{r}})] \\
&= \mathrm{i}\hbar \left[\frac{\partial \Psi_0}{\partial t}(1+(u+iv)e^{i\mathbf{k}\cdot\mathbf{r}}) + \Psi_0\left(\frac{\partial u}{\partial t} + i\frac{\partial v}{\partial t}\right)e^{i\mathbf{k}\cdot\mathbf{r}}\right].
\end{align}
Using $\mathrm{i}\hbar \frac{\partial \Psi_0}{\partial t} = \mu\Psi_0$, we obtain:
\begin{equation}
\mathrm{i}\hbar \frac{\partial \Psi}{\partial t} = \mu\Psi_0(1+(u+iv)e^{i\mathbf{k}\cdot\mathbf{r}}) + \mathrm{i}\hbar\Psi_0\frac{\partial u}{\partial t}e^{i\mathbf{k}\cdot\mathbf{r}} - \hbar\Psi_0\frac{\partial v}{\partial t}e^{i\mathbf{k}\cdot\mathbf{r}}.
\end{equation}
The kinetic energy term under the local density approximation ($k\sigma \gg 1$) becomes:
\begin{align}
-\frac{\hbar^2}{2m}\nabla^2\Psi &\approx -\frac{\hbar^2}{2m}[\nabla^2\Psi_0(1+(u+iv)e^{i\mathbf{k}\cdot\mathbf{r}}) - k^2\Psi_0(u+iv)e^{i\mathbf{k}\cdot\mathbf{r}}].
\end{align}
The interaction term to first order in perturbation is:
\begin{align}
g(t)|\Psi|^2\Psi &\approx g(t)\bar{n}_0\Psi_0[1 + 2(u+iv)e^{i\mathbf{k}\cdot\mathbf{r}} + (u-iv)e^{-i\mathbf{k}\cdot\mathbf{r}}].
\end{align}
The gravitational potential term is:
\begin{align}
m\Phi\Psi &= m(\Phi_0 + \delta\Phi)\Psi_0(1+(u+iv)e^{i\mathbf{k}\cdot\mathbf{r}}) \\
&= m\Phi_0\Psi_0(1+(u+iv)e^{i\mathbf{k}\cdot\mathbf{r}}) + m\delta\Phi\Psi_0,
\end{align}
with the gravitational potential perturbation given by:
\begin{equation}
\delta\Phi = -\frac{8\pi G m \bar{n}_0 u}{k^2}\cos(\mathbf{k}\cdot\mathbf{r}) = -\frac{4\pi G m \bar{n}_0 u}{k^2}(e^{i\mathbf{k}\cdot\mathbf{r}} + e^{-i\mathbf{k}\cdot\mathbf{r}}).
\end{equation}
Subtracting the equilibrium equation and keeping only terms proportional to $e^{i\mathbf{k}\cdot\mathbf{r}}$ (resonant terms), we obtain after dividing by $\Psi_0e^{i\mathbf{k}\cdot\mathbf{r}}$:
\begin{align}
\mu(u+iv) + \mathrm{i}\hbar\frac{\partial u}{\partial t} - \hbar\frac{\partial v}{\partial t} &= 
\frac{\hbar^2 k^2}{2m}(u+iv) + 2g(t)\bar{n}_0(u+iv) + m\Phi_0(u+iv) \\
&\quad - \frac{4\pi G m^2 \bar{n}_0 u}{k^2} - i\hbar\gamma(u+iv).
\end{align}
Separating real and imaginary parts yields the coupled equations:
\begin{subequations}
\begin{align}
-\hbar\frac{\partial v}{\partial t} &= \frac{\hbar^2 k^2}{2m}u - \mu u + 2g(t)\bar{n}_0 u + m\Phi_0 u - \frac{4\pi G m^2 \bar{n}_0 u}{k^2} + \hbar\gamma v, \\
\hbar\frac{\partial u}{\partial t} &= \frac{\hbar^2 k^2}{2m}v - \mu v + 2g(t)\bar{n}_0 v + m\Phi_0 v - \hbar\gamma u.
\end{align}
\end{subequations}
Using the equilibrium relation $\mu = g_0\bar{n}_0 + m\Phi_0$ and applying spatial averaging over the Gaussian background, we obtain Eqs.~(\ref{realm_eq}) and (\ref{imagm_eq}) in the main text.

\subsection{Reduction to Second-Order Equation}

From Eq.~(\ref{imagm_eq}), we solve for $v$:
\begin{equation}
v = \frac{\hbar\frac{d u}{d t} + \hbar\gamma u}{\frac{\hbar^2 k^2}{2m} + [2g_0\alpha\cos(\omega t) + g_0]\bar{n}_0}.
\end{equation}
Defining $A = \frac{\hbar^2 k^2}{2m} + [g_0 + 2g_0\alpha\cos(\omega t)]\bar{n}_0$, this simplifies to:
\begin{equation}
v = \frac{\hbar}{A}\left(\frac{d u}{d t} + \gamma u\right). \label{v_expression}
\end{equation}
Differentiating with respect to time:
\begin{equation}
\frac{d v}{d t} = \hbar\left[-\frac{1}{A^2}\frac{d A}{d t}\left(\frac{d u}{d t} + \gamma u\right) + \frac{1}{A}\left(\frac{d^2 u}{d t^2} + \gamma\frac{d u}{d t}\right)\right], \label{dvdt}
\end{equation}
where $\frac{d A}{d t} = -2g_0\alpha\bar{n}_0\omega\sin(\omega t)$.
Substituting Eqs.~(\ref{v_expression}) and (\ref{dvdt}) into Eq.~(\ref{realm_eq}) and applying the adiabatic approximation (neglecting $\frac{d A}{d t}$ terms when $\omega \ll \Omega_k$), we obtain after simplification:
\begin{equation}
\frac{d^2 u}{d t^2} + 2\gamma\frac{d u}{d t} + \left[\frac{\hbar^2 k^4}{4m^2} + \frac{g(t)\bar{n}_0 k^2}{m} - \Omega_J^2\right]u = 0,
\end{equation}
which is Eq.~(\ref{second_order}) in the main text.

\end{appendix}
\bibliography{refs}
\bibliographystyle{elsarticle-num-names}

\end{document}